# A New Trust Reputation System for E-Commerce Applications

Hasnae RAHIMI[1] and Hanan EL BAKKALI[2]
Information Security Research Team (ISeRT), University Mohamed V-Souissi, ENSIAS
RABAT, Morocco

**Abstract**

Robust Trust Reputation Systems (TRS) provide a most trustful reputation score for a specific product or service so as to support relying parties taking the right decision while interacting with an e-commerce application. Thus, TRS must rely on an appropriate architecture and suitable algorithms that are able to improve the selection, storage, generation and classification of textual feedbacks.

In this work, we propose a new architecture for TRS in e-commerce applications. In fact, we propose an intelligent layer which displays to each feedback provider, who has already given his recommendation on a product, a collection of prefabricated feedbacks related to the same product. Our main contribution in this paper is a Reputation algorithm which studies the user's attitude toward this selection of prefabricated feedbacks. As a result of this study, the reputation algorithm generates better trust degree of the user, trust degree of the feedback and a better global reputation score of the product.

**Keywords**: *E-Commerce, Trust management, Reputation Trust System, Textual feedbacks.*

## 1. Introduction

Trust is an important factor in any social relationship and especially in commerce transactions. In the e-commerce context, there is a lack of direct trust assessment. Although cryptography, electronic signatures and certificates assist users in order to make the transaction more secure, they remain insufficient to construct a trustful reputation about a product or a service [1]. As a result, users are not able to conceive a reputation for the product without any additional help [2].

In such circumstances, Trust Reputation Systems (TRS) are solicited in e-commerce applications so as to create trustworthiness, among a group of participants, toward transactions' circumstances, products' characteristics and toward users' passed experiences. In fact, e-commerce users prefer to focus on users' opinions about a product, in order to conceive their own trust and reputation experience. Users believe in their common interest which is to know about the trustworthiness of the transaction and product [3,4]. Therefore feedbacks or reviews, scores, recommendations and any other information given by users are very important for the trust reputation assessment. However, the reliability of this information needs to be verified.

TRS are indeed essential mechanisms that aim to detect malicious interventions of users whose intention is to falsify the Reputation score of a product positively or negatively.

In the literature, there are many works such as [2, 3, 4, 5] that propose algorithms for calculating a reputation or defining a specific set of possible reputations or ratings. However, few of them such as [6,7,8] have been devoted to the semantic analysis of textual feedbacks in order to generate a most trustful trust degree of the user.

In contrast to these papers, we analyse the attitude adopted by the user toward specific prefabricated textual feedbacks. This selection of reviews is fabricated thanks to a text mining algorithm which is not detailed in this paper. In fact, the user is going to give his opinion (like/dislike) on those prefabricated feedbacks. Each prefabricated feedback has a degree of trustworthiness. However, hypotheses concerning the text mining algorithm are analysed in term of availability and realization. In fact, the text mining algorithm is supposed to classify users' feedbacks by categories in a knowledge base depending on their semantic content. The text mining algorithm is supposed to verify also the concordance between the user's appreciation on a product and the review associated to it.

The knowledge base needs a learning algorithm which is also going to be detailed in a future work. However we give in this paper a brief description of the inputs of the learning algorithm.

In this paper, our main contribution is a reputation algorithm which uses the selection of prefabricated feedbacks in order to analyse the user's attitude and intention toward the product. According to the user's opinion and to the trustworthiness degree of each feedback, the proposed reputation algorithm generates a better trust degree of the user. The algorithm generates also a most trustful reputation score of a product using the trust degree of the user as a coefficient. At the end of the execution, the algorithm applies a trustworthiness degree to the feedback.

The remainder of this paper is structured as follows. In section 2, we remind the terminology of trust and reputation systems. Section 3 presents some related work. The architecture of the TRS and the hypotheses related to the text mining algorithm are explained in section 4. In the same section, we analyse and present our proposed Reputation algorithm. Finally, we come up with some concluding remarks and an outlook on the future work.

## 2. Trust and Reputation Background

Concepts of 'Trust' and 'Trustworthiness' can seem to be well understood, but, in reality there is no common agreeing on what they precisely mean, on how to 'calculate' and use them. Even in the context of e-commerce security, we find number of definitions. We present hereafter some of them.

## 2.1 Trust and Reputation definitions

Among the available trust definitions that we could find in the literature, we choose the following to highlight the main features of trust and related concepts and we will refer to them in the following statements:

Definition 1: Trust is the firm belief in the competence of an entity to act dependably, reliably and securely within a specific context [9].

Definition 2: Trust is also defined as the ability to rely on someone or something, to rely on its truthiness, on its strength to prove its reliability. In e-commerce, being trustful (or trustworthy) is a quality characterizing a product that a user claims to know either intuitively or from a personal past experience which is more trustful or other users' experiences [3, 10].

Thus, in e-commerce applications, trust which is not based on logical evidence corroborated by users real experiences and analytical examination is useless and doesn't help generating a reliable reputation that allow propagating this trust among other users. Let's note here, that users' experiences are generally represented by rating and semantic feedbacks.

We use this definition in order to define the user's trust degree. In fact, the user's trust degree represents the degree of trust related to the user. We define also the feedback's trustworthiness which is a degree of trustworthiness related to the review. The product has also a trust score which refers to the trustworthiness of the product.

Definition of Reputation: the concept of reputation is closely linked to that of trust and trustworthiness. Indeed, Reputation is generally said or believed to be about a person's or things' character or standing which is a real proof of subjectivity [10, 11].

We will refer to this definition of reputation in the following statements.

## 2.2 Trust Reputation System definitions

Definition 1 Reputation systems are one of the established mechanisms to assist consumers in making decision in online shopping [1]. They allow e-commerce participants to evaluate the reputation of a product, a transaction, an online merchant according to their own experience or other users' one [12,13].

As a result, TRS help people detect trustworthy parties and influence buyers that may base their buying decision on the past experiences of other participants [4].

Definition 2 TRS are also defined as trustworthiness providers that assess trust among a community of users as an option to help users identify and detect reliable relationships in the Internet. Indeed, the purpose of Trust Reputation Systems is to allow parties to rate each other [10]. In fact, they focus on providing consumers the most trustful reputation, represented by a rating or/and a textual feedback, which intends to convince users to whether rely on the product or not. This user's trust is based on the probable trust they have on the provider of the rating and the feedback [3,14,15].

In this paper, we will use this definition of TRS in the following statements.

## 3. Related works

Many works such as [2, 3, 4, 5] propose TRS architectures together with different algorithms to calculate the reputation score related to a product. Nevertheless, few research works on TRS has considered the semantic analysis of feedbacks and especially the trust degree of the user in the calculus of products' trust scores.

Even in studies attempting to provide more complex reputation methods such as [6,10, 13,16,17], some issues are still not taken into consideration, such as the inclusion of the trust degree of the user in the calculus of a trustful reputation score for a product, the update of the trust degree of the user "at any intervention", the freshness of the rating and especially the feedback, the concordance between the given rating which is a scalar value and the textual review associated to it.

Unlike those TRS, our proposed design treats these issues and uses a reputation algorithm that includes semantic analysis of textual feedbacks in order to calculate the trust degree of the user. This proposed reputation algorithm calculates also the global reputation score of the product using the trust degree of the user as a coefficient.

For example, the authors of [2] propose a method that uses subjective logic in order to analyse trust network (TNA-SL). Hence, this method aims to model in a simple way the relationship between different agents. A single arc means a single trust relationship between two nodes A and B [A;B] meaning that A trusts B. However, this trust should have degrees that can represent how much A trusts B. This issue is not taken into account in the paper [2]. However we should calculate the trust degree of the arc and also the trust degree of the nodes.

In the proposed architecture, for each user who wants to leave a rating (appreciation) and a textual feedback (semantic review), we analyse his attitude toward a number of short and selected feedbacks prefabricated and stored by product in the knowledge base. This user's review is going to be reached by any other user. Then, we suppose that we have a path relaying all the users (the nodes). Any feedback can be an arc between 2 nodes or more. As a result, we need to know the trust degree of the user and determinate the trust degree of the feedback.

Another factor which is important in the analysis realized by a TRS is the date of the creation or the establishment of the arc (the freshness of the arc). The most recent arc, which relays two nodes having the same interest on a topic or a product, is more meaningful and useful than an old one. In fact, the 'feedback freshness' issue is a very neglected and important issue. Consequently, a part of our contribution is to take into consideration the freshness of the feedback in our reputation algorithm.

Besides, the authors of [7] use an approach that calculates the trust weight. In fact, once the transaction is carried out between the Web Service Providers WSP and the Web Service Consumers, a reward or a punishment is applied to users and WSPs according to the accuracy and reliability of their recommendations.

A simple mechanism will be established to measure the divergence between the final satisfaction of the user and the previously given recommendation of users. As a result, the authors of [3] and [7] focus on the recommendation and the satisfaction, which are both a subjective feedback, in order to reward or punish a user. However, we must not rely on the user's satisfaction about a given recommendation because it could be falsified if the user is ill-intentioned.

In this paper, we do not measure the divergence between the user's satisfaction and the users' recommendations. But we establish an algorithm that analyses the divergence between his first given recommendation and his opinion on a selection of short reviews. The algorithm rewards users virtual credits which represent their trust degree according to the review's trustworthiness and to the user's opinion (either a like or a dislike) on this feedback. In fact, the approach represents basically a verification of the interaction party as a human by inviting him to like/dislike a very short selection of user-based feedbacks. Performing the like/dislike task is not a time consuming procedure and is necessary to validate the recommendation of the informed user. In e-commerce applications, we are used to fill out forms in order to validate or get a request. Then this approach seems to be very suitable for e-commerce applications.

## 4. Requirements of proposed Trust Reputation System

### 4.1 Trust Reputation System Architecture

At the beginning, the user gives an appreciation (rating) and a textual feedback on a specific product. The TRS need a text mining algorithm which aims to get the given information and verify the concordance between the user's given appreciation and the textual feedback, so as to avoid and eliminate any contradiction.

Once the concordance verified, we redirect the user to an interface of selected pre-fabricated feedbacks. So as long as we add feedbacks in the data base of origin, a text mining algorithm is going to make pre-fabricated feedbacks with different categories and fill out the knowledge base (Fig. 1 shows the architecture). The text mining algorithm would contain a part of learning in order to automatically fill out the knowledge base. The user is invited to like or dislike each feedback of the dis-played selection. Each feedback has already a degree of trustworthiness which represents the trust degree of the user who is the provider of the feedback. The user can choose the number of short feedbacks like and dislike (min=4 and max=10).

Then the proposed reputation algorithm gets the user's opinion on each review (like/dislike) in addition to the trustworthiness degree of the liked/disliked feedback and uses them to generate a trust degree for the user.

The architecture hereafter represents the connection between the e-commerce application and the solicited TRS showing the intervention of both the text mining and the Reputation algorithm.

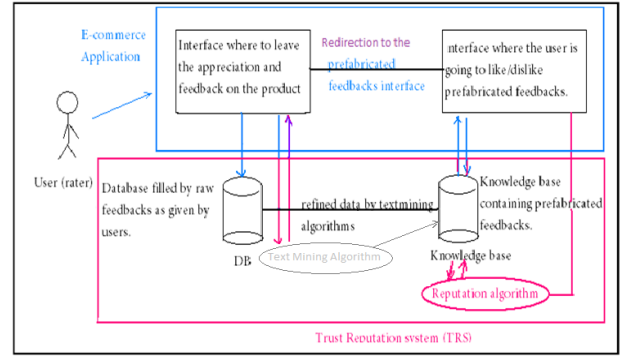

Fig. 1. Trust Reputation System Architecture

A product can be replaced by a service or even a participant. This analysis can be applied on many types of products, services and agents. But, in this paper, we're analysing the case of a product.

*a. Hypotheses of our approach*

Our TRS architecture is based on 2 algorithms: the text mining algorithm and the reputation algorithm. On one hand, the TRS relies on semantic feedbacks' analysis realized by a text mining algorithm in order to verify the concordance between the appreciation and the textual feedbacks. Moreover, we need a text mining algorithm in order to classify the feedbacks by categories in a knowledge base.

On the other hand, we use the proposed reputation algorithm to generate the user's trust degree and the trustworthiness degree of the product using his trust degree.

However in this work, we focus on the reputation algorithm and we suppose that the following hypotheses concerning the text mining algorithm are verified.

### 4.2 Hypotheses about the knowledge base

As seen before in the TRS architecture, we need a knowledge base where we can store feedbacks pre-fabricated using users' feedbacks and a text mining algorithm. Every e-commerce application provide a vast source of information accessible to users, but understandable only to humans. Then, the objective of the knowledge base associated to the text mining algorithm is to automatically collect for each product a number of characteristics and properties which are going to help analyse the meaning of each feedback. Such a knowledge base would help classify feedbacks by their semantic content, products' categories and properties. Consequently, the knowledge base would enable much more effective retrieval of feedbacks by products' categories etc. Thanks to a text mining algorithm, we will fill the knowledge base with pre-fabricated feedbacks with different types summarizing other users' feedbacks. We can even choose users' feedbacks which are already summarized and store them directly in the knowledge base.

In a future work, we aim to develop a trainable information extraction system that takes several inputs. We can suppose that an ontology that defines the properties by products' categories (e.g., camera, screen, weight… etc for phones, processor, hard drive… etc for laptops) is an interesting input. Furthermore, this ontology must be up to date and refreshed for any new product or changes on the existing products.

To evoke an absent property in the feedback doesn't lead always to a contradiction. The user can talk about the absence of some properties which are important for him. For example, if the user writes that the negative thing about this hotel is that there is no swimming pool and the hotel does not have one. He's telling the truth. Then obviously the presence of an absent property in a feedback doesn't mean that there is a contradiction. The text mining algorithm is supposed to clarify as much as possible these assumptions. That's why we suppose that a set of training data consisting of subjective expressions that represent instances of users' opinions is an important input. Given these inputs, the system learns to extract information from pages and hyperlinks of any e-commerce application that applies our TRS.

### 4.3 Availability of text mining functions to be used in a future work

Many text mining algorithms already exist and can eventually serve our requests such as [18, 19, 20, 21]. Most of those existing algorithms or tools provide functions that can classify textual feedbacks by analysing their content and store them by categories in a knowledge base.

Automap

For instance, the authors of [18] propose a frequent pattern mining algorithm to mine a bunch of reviews and extract product features. Their experimental results indicate that the algorithm outperforms the old pattern mining techniques used by previous researchers of mining algorithms. Then this mining algorithm could be very useful to extract product feature from our textual feedback. However we need the sentiment analysis for each statement, of the textual feedback, containing the product feature in order to classify the feedback. For example, the following sentence shows a negative feedback on a cellphone: "It is not easy to carry." "Weight" is an implicit feature of the cellphone which is implied from the sentence. In fact, the mining algorithm proposed in [18] can extract the weight as a feature from the sentence below. The opinion expressed about the feature is negative then we have to classify it as a negative feedback. The sentiment analysis model proposed in [19] takes a collection of reviews as input, and processes them using three core steps, Data Preparation, Review Analysis and Sentiment Classification. The results produced by such a model are the classifications of the reviews, the evaluative sentences, or opinions expressed in the reviews. Consequently, we can use this sentiment analysis model to classify the textual feedbacks.

The authors of [20] present two methods for determining the sentiment expressed by a movie review (textual feedback). They examine the effect of valence shifters (negations, intensifiers and diminishers) on classifying the reviews. The first method classifies reviews based on the number of positive and negative terms they contain. The second method uses a Machine Learning algorithm, Support Vector Machines. The authors show that combining the two methods achieves better results than using each method alone.

We can also use this combination of methods in order to classify our textual feedbacks in their different categories.

The research done in [21] presents a text mining package which provides a framework for text mining applications within R. They present techniques for count-based analysis methods, text clustering text classification and String Kernels. This TM package in R can be used to respond to our text mining purposes in a future work.

In the following statements, we will discuss the aim and the use of the text mining functions and which are going to be detailed in a future work.

### 4.4 Function verifying the concordance between the appreciation and the textual feedback

Before redirecting the user to an interface containing pre-fabricated feedbacks, we are going to verify first of all the concordance between the user's rating (appreciation) and his textual feedbacks in order to avoid any contradiction. For that purpose, many text mining functions which analyse a textual review and a score (appreciation) exist already and we can appropriately adapt it to our purpose in a future work.

This verification aims to insure that the appreciation is reflecting the content of the feedback and vice versa.

For that reason, we suppose that we have a text mining function named "Test_concordance" which takes two parameters: the first one is the appreciation as a float and the second one is the textual feedback as a string. It returns a Boolean value: True if there is a concordance and False if not. We can present its prototype as follows:

```
Boolean  Test_concordance (float Appreciation, String Feedback)
```

### 4.5 Function classifying feedbacks

Furthermore, we suppose that our text mining algorithm is using a function of feedbacks' classification which is based on one of the existing text mining algorithms and which will be eventually adapted appropriately to our TRS architecture in a future work.

The selection of pre-fabricated feedbacks from different types is going to be liked or disliked by the user according to his experience and intention.

In that way, we can analyse the user's intention behind his intervention on the e-commerce application, using other pre-fabricated feedbacks with different types.

However, in order to obtain this selection of different types of feedbacks, we need a text mining function that we can call "classfeed" which examines the content of users' feedbacks and classify them in four different categories in the knowledge base.

One of the chosen Text mining algorithm or tool would need inputs briefly discussed in (section 4.2.1). Furthermore, our reputation algorithm is going to apply to each feedback a trustworthiness degree in the threshold [-10,10]. The closest is the trustworthiness degree to 10, the most trustworthy the feedback is. The closest is the trustworthiness degree to -10, the very untrustworthy is the feedback.

If the feedback is relatively trustworthy, its score would be in ]0,10] else it would be included in [-10,0].

The use of this threshold is very interesting since it is large but not excessively. Then we can have a large set of values to determinate in a better way the trustworthiness of a feedback. We describe the feedbacks categories as follows:

Positive feedbacks: represent opinions expressing a positive point of view about the product. Those ameliorative opinions contain a positive content concerning the product. Then, the adjective positive is referring to the nature of the content of the feedback not its trustworthiness.

However, each feedback whatever is its type can have either a positive trustworthiness or a negative trustworthiness.

Negative feedbacks: represent opinions talking negatively about the product.

Mitigated feedbacks: represent feedbacks that are talking positively about some aspects of the product and negatively about other aspects.

Contradictious feedbacks: represent feedbacks with a contradictious content. A malicious program can be the source of such feedbacks.

Concerning, the rating of this type of feedback, each contradictory feedback would have -10 as a trustworthiness degree.

## 5. Proposed reputation algorithm

### 5.1 General overview of the TRS

Before giving details on the approach of the Reputation algorithm, we will start first with giving an overview on the steps of the algorithm:

1. Verify the concordance between the appreciation and the textual feedback.
2. Display to the user a selection of the most recent pre-fabricated feedbacks with different types (freshness of feedbacks), if the concordance is verified. This selection of feedbacks is to be liked or dis-liked by the user.
3. Extract data from data base concerning the trustworthiness of the liked or disliked feedback and the trust degree of the user.
4. Generate / update the trust degree of the user using the trustworthiness of the feedback and the user's choice (like/dislike).
5. Standardize the trust degree of the user in order to respect the threshold [-10,10].
6. Generate the global trust score of the product using the user's trust degree as a coefficient.

### 5.2 Concordance between the appreciation and the textual feedback step

The algorithm hereafter describes the first step of the reputation algorithm.

```
Boolean concordance;
concordance =Test_ concordance (float appreciation, string feedback) ;
If (concordance)
    URL (url_feedbacks_interface);
    //redirection to the pre-fabricated feedbacks interface
Else
    URL (url_page);
    // we thank the user for his intervention and we put
    // him temporally in a blacklist for unconformity
```

After verifying the concordance between the appreciation and the textual feedback, we're going to redirect the user to the selection of pre-fabricated feedbacks.

### 5.3 Selection of fresh pre-fabricated feedbacks step

In the redirected platform displayed after the user's request to validate his given information, we need to select for the user some prefabricated feedbacks related to the product with a most Recent Date. In fact, the information about the reputation is more reliable when the date is more recent, because the user for instance can live another experience that changes his opinion about the product and even with time a product can remain untrustworthy after being trustworthy and vice versa. Then, the freshness of a feedback is very important since the reputation of a product can change with the time factor.

As a result, we can select different type of feedbacks from their table in the Knowledge base; group them by the specific product having the recent Date.

### 5.4 Extraction of useful data step

The function "get-infos-click" gets some information in order to calculate the trust degree of the user. The function gets also the previous trust degree of the user if he has already given a rating in the application for instance. The user choices either "like" or "dislike" is an important parameter to determine his trustworthiness. The trustworthiness of the feedback is also needed.

```
Function get-infos-click (int idfeedback) as list
{ double Feedtrustworth;
  //the feedback trustworthiness stored in knowledg base
  // its value is between -10 and 10

  String Userchoice;
  // represents the user's choice either it is a "like" or a "dislike"

  String login=get_user_login();
  /*to get the user login in order to get his trust degree if he had
      already done an intervention in the application*/

  Feedtrustworth= getfeedtrustworth (idfeedback);
  /*this function gets the trustworthiness of the feedback either
  positive or negative value between -10 and 10*/

  Double degree_trust_user =get_trust_degree_user (login);
  Userchoice=getuserchoice (idfeedback);

  // this function get the user choice after the click from
  the interface

  List listinfos=[ Feedtrustworth, Userchoice, degree_trust_user];
  Return listinfos;
}
```

### 5.5 Generation/update of the user's trust degree step

After getting the parameters going to be used in the next calculus, we're going to calculate the trust degree of the user taking into consideration the value of the trustworthiness of the feedback, the user's choice made on this feedback and his previous trust degree calculated in the previous intervention. The calculus of the trust degree of the user can be an update if the user has already a trust degree.

Our proposed algorithm rewards the user by incrementing his trust degree if he likes a trustworthy feedback or he dislikes an untrustworthy one. The algorithm also punishes the user if he likes an untrustworthy feedback or dislikes a trustworthy one. When the user choice is a "like", the greatest is the feedback's trustworthiness, the greatest the reward would be and vice versa. And when the user dislikes a feedback, the greatest is the untrustworthiness of the feedback, the greatest the reward would be and vice versa.

We consider that all users in their first participation in the application are neutral and have the same initial trust degree which is 0/10. The following function generates the trust degree of the user by rewarding or/and punishing him:

```
function calculate_degree_trust_user () as double
{
list listinfos;
Int idfeedback=get_idfeedback();
Double Ufeedtrustworth;
Listinfos=get_infos_click (idfeedback);
Double Feedtrustworth= Listinfos[0];
String Userchoice= listinfos[1];
double Degree_trust_user=Listinfos[2];
/*with the value of 0 at the first intervention*/
{ // according to the applicable case:
    Do:
    Case 1: (0<feedtrustworth<=3) and (userchoice="like")
       Or (-3=<feedtrustworth<=0) and (userchoice="dislike")

    Do:      Degree_trust_user+=0.25
    Case 2: (3<feedtrustworth<=5) and (userchoice="like")
       Or (-5=<feedtrustworth<-3) and (userchoice="dislike")

    Do:  Degree_trust_user+=0.5
    Case 3: (5<feedtrustworth<=7) and (userchoice="like")
       Or (-7=<feedtrustworth<-5) and (userchoice="dislike")

    Do:  Degree_trust_user+=0.75
    Case 4: (7<feedtrustworth<=8) and (userchoice="like")
       Or (-8=<feedtrustworth<-7) and (userchoice="dislike")

    Do:  Degree_trust_user+=1
    Case 5: (8<feedtrustworth<=9) and (userchoice="like")
       Or (-9=<feedtrustworth<-8) and (userchoice="dislike")

    Do:  Degree_trust_user+=1.5
    Case 6: (9<feedtrustworth<=10) and (userchoice="like")
       Or (-10<feedtrustworth<-9) and(userchoice="dislike")

     Do: Degree_trust_user+=2
    Case 7: (-3=<feedtrustworth<=0) and (userchoice="like")
       Or (0<feedtrustworth<=3) and (userchoice="dislike")

    Do:      Degree_trust_user-=0.25
    Case 8: (-5=<feedtrustworth<-3) and (userchoice="like")
       Or (3<feedtrustworth<=5) and (userchoice="dislike")

    Do      Degree_trust_user-=0.5
    Case 9: (-7=<feedtrustworth<-5) and (userchoice="like")
       Or (5<feedtrustworth<=7) and (userchoice="dislike")

    Do      Degree_trust_user-=0.75
    Case 10: (-8=<feedtrustworth<-7) and (userchoice="like")
       Or (7<feedtrustworth<=8) and (userchoice="dislike")

    Do      Degree_trust_user-=1
    Case 11: (-9=<feedtrustworth<-8) and (userchoice="like")
       Or (8<feedtrustworth<=9) and (userchoice="dislike")

    Do      Degree_trust_user-=1.5
    Case 12: (-10<feedtrustworth<-9) and (userchoice="like")
       Or (9<feedtrustworth<=10) and (userchoice="dislike")

    Do      Degree_trust_user-=2
    Case* 13: (feedtrustworth=-10) and (userchoice="like")
    /*a   contradictory feedback*/
    Do      Degree_trust_user=-10
    }
}
```

## 5.6 Standardize the trust degree of the user step

The following algorithm aims to respect the threshold [-10,10]. It applies the trust degree of the user to his feedbacks' trustworthiness.

```
// to respect the threshold [-10;10]
If (Degree_trust_user<-10)
    Degree_trust_user=-10;

Else if (Degree_trust_user>10)
    Degree_trust_user=10;
Return degree_trust_user;
}
// the end of the function

 /* apply the trust degree of the user to the degree
of trustworthiness of his feedback */

Ufeedtrustworth=Degree_trust_user;
```

The function returns the trust degree of the user updated according to his current participation. As a result, if his trust degree is positive we will take into account his given appreciation. However, if his trust degree is negative, we will not include his appreciation in the calculus of the global trust score of the product and we can preserve his feedback in order to use it to fabricate other feedbacks. Then his feedback would be considered as untrustworthy as his provider and vice versa. Consequently, we apply the trust degree of the user to the degree of trustworthiness of the user's feedback as shown in the last line of the pseudo-code below.

## 5.7 Calculus of the global trust score of the product using the user's trust degree step

After that, we have to generate the global trust reputation score of the product using the user's appreciation (rating) and his trust degree. In fact, a possible example for such rating method might be school marks and coefficients. Actually, at school, when a course is important for a certain field, its coefficient would be great and then the effect of its mark would be greater. In the same context, we consider the trust degree of the user as a coefficient and his appreciation as a mark. Consequently, to calculate the global trust score of the product, we sum all the appreciation values multiplied by their respective coefficient and then divide the result of the summation on the summation of all coefficients:

$$\frac{X + bY}{a + b} = \frac{\sum_{1st\ user}^{last\ user} Appreciation * trust\ degree}{\sum_{1st\ user}^{last\ user} trust\ degree}$$

* "X" represents the summation of all users' appreciations.
*"Y" represents the new appreciation given by the user.
*"b" represents the new coefficient to be added, and "a" represents the summation of all users' trust degrees.

We can store the "X" and the "a" in different areas so we can get them separately and then calculate easily: $\frac{X+bY}{a+b}$

As a result, we update the global trust score of the product. All in all, our proposed algorithm aims to calculate the trust degree of the user according to his adopted attitude toward some fake and prefabricated feedbacks related his targeted product.

## 6. Conclusion and Future work

In this paper, we design a Trust Reputation System based on the analysis of the user's attitude toward a collection of prefabricated textual feedbacks. We propose a Reputation algorithm attempting to calculate the trust degree of the user according to his subjective choice either "like" or "dislike" and according to the feedback trustworthiness. The proposed reputation algorithm calculates also the global trust reputation score of the product and generates the trustworthiness of the user's given feedback.

In this work, we give some hypotheses concerning a text mining algorithm which is supposed to classify users' feedbacks by categories in a knowledge base and verify the concordance between the given appreciation and the feedback associated to it.

As a perspective, we will relieve these assumptions in our experimental analysis to more extensively evaluate the effectiveness, the robustness and the improvement contribution of our Trust Reputation System.

## References

[1] Anna Gutowska and Andrew Sloane: Modelling the B2C Marketplace: Evaluation of a Reputation Metric for e-commerce. Proceedings of Web Information Systems and Technologies - WEBIST , pp. 212-226, 2009.
[2] Audun Jøsang Ross Hayward Simon Pope: Trust Network Analysis with Subjective Logic. Proceedings of the Second International Conference on Emerging Security Information, Systems and Technologies (SECURWARE 2008), Cap Esterel, France, August 2008.
[3] Yuan Liu, Jie Zhang, and Quanyan Zhu: Design of a Reputation System based on Dynamic Coalition Formation. Proceedings of Third International Conference, SocInfo 2011, Singapore, October 6-8, 2011. Proceedingspp 135-144.
[4] Jennifer Golbeck James Hendler: Inferring Reputation on the Semantic Web. In the proceedings of WWW 2004, May 17-22, 2004, New York, NY USA. ACM.
[5] Fereshteh Ghazizadeh Ehsaei, Ab. Razak Che Hussin: Acceptance of Feedbacks in Reputation Systems: The Role of Online Social Interactions Information Management and Business Review Vol. 4, No. 7, pp. 391-401, July 2012 (ISSN 2220-3796).
[6] Audun Jøsang, Jennifer Golbeck: Challenges for Robust Trust and Reputation Systems. 5th International Workshop on Security and Trust Management (STM 2009), Saint Malo, France, September 2009.
[7] Félix Gómez Mármol, Joao Girao, Gregorio Martínez Pérez: TRIMS, a privacy-aware trust and reputation model for identity management systems. In the proceeding of Computer Networks, 54(16):2899-2912, September 2010.
[8] Jennifer Golbeck James Hendler: Inferring Reputation on the Semantic Web. In the proceedings of WWW 2004, May 17-22, 2004, New York, NY USA. ACM.
[9] A. Jøsang (1996). The right type of trust for distributed systems. In C. Meadows, editor, Proc. of the New Security Paradigms Workshop. ACM, 1996.
[10] Josang, Audun and Ismail, Roslan and Boyd, Colin. A survey of trust and reputation systems for online service provision. Decision Support Systems 43(2):pp. 618-644. (2007)
[11] Roger C. Mayer, James H. Davis and F. David Schoorman. An Integrative Model of Organizational Trust The Academy of Management Review.Vol. 20, No. 3 (Jul., 1995), pp. 709-734.
[12] Hasnae Rahimi, El bakkali hanan: Toward a New Design of Trust Reputation System in e-commerce,. In the proceedings of ICMCS (International conference on Multimedia computing and systems, Tangier, Morocco, IEEE 2012).
[13] Huanyu Zhao and Xin Yang Xiaolin Li: WIM: A Wage-based Incentive Mechanism for Reinforcing Truthful Feedbacks in Reputation Systems. In the Proceedings of the Global Communications Conference, 2010. GLOBECOM 2010, 6-10 December 2010, Miami, Florida, USA. IEEE 2010.
[14] Sherrie Komiak: The Effects of Perceived Information Quality and Perceived System Quality on Trust and Adoption of Online Reputation Systems. AMCIS 2010 Proceedings.
[15] Audun Jøsang, Walter Quattrociocchi, and Dino Karabeg: Taste and Trust. In the Proceedings of IFIPTM 2011 Copenhagen, June/July 2011.
[16] Ray and S. Chakraborty, "A Vector Model of Trust for Developing Trustworthy Systems", Proc. European symposium on Research in Computer Security, Springer-Verlag, pp. 260-275, 2004.
[17] Sandra Steinbrecher, Stephan Groß, and Markus Meichau: Jason: A Scalable Reputation System for the semantic Web. In the proceedings of IFIP International Federation for Information Processing 2009.
[18] Roliana Ibrahim, Shirin Noekhah and Niloufar Salehi Dastjerdi: A Frequent Pattern Mining Algorithm for Extraction of Customer Reviews. IJCSI International Journal of Computer Science Issues, Vol. 9, Issue 4, No 1, July 2012.
[19] Cane W. K. Leung and Stephen C. F. Chan. J. Wang (Eds.): Sentiment Analysis of Product Reviews. , Encyclopedia of Data Warehousing and Mining - 2nd Edition, Information Science Reference, pp. 1794-1799, August 2008.
[20] Alistair Kennedy and Diana Inkpen: Sentiment Classification of Movie Reviews Using Contextual Valence. Digital Libraries: Universal and Ubiquitous Access to Information Lecture Notes in Computer Science Volume 5362, 2008, pp 184-193.
[21] Ingo Feinerer, Kurt Hornik, and David Meyer. Text mining infrastructure in R. Journal of Statistical Software, 25(5):1-54, 2008.